\documentstyle[twocolumn,aps,epsfig]{revtex}
\begin{document}
\draft
\title{ Montecarlo simulation of the role of defects as a  melting mechanism} 
\author{L. G\'omez$^{1,2}$, A. Dobry$^{1,2}$ and H. T. Diep$^1$} 
\address{$^1$ Departamento de F\'{\i}sica, Universidad Nacional de Rosario, \\
and Instituto de F\'{\i}sica Rosario, Avenida Pellegrini 250,\\
2000 Rosario, Argentina\\ and\\
$^2$Laboratoire de Physique Th\'eorique et Mod\'elisation,
Universit\'e de Cergy-Pontoise,\\
5, mail Gay-Lussac, Neuville sur Oise, 95031 Cergy-Pontoise Cedex, France\\
E-mail: diep@ptm.u-cergy.fr}
\maketitle
\begin{abstract}
We study in this paper the melting transition of
a crystal of fcc structure with the Lennard-Jones potential
, by using isobaric-isothermal Monte Carlo simulations.
 Local and collective updates
are sequentially used to optimize the convergence.  
We show the important role played by defects
in the melting mechanism in favor of modern
melting theories.
\end{abstract}
\vspace{1cm}
%\newpage

\section{Introduction}
Melting of crystals has always been an exciting subject in condensed matter
physics\cite{Aschroft}.  The Lindemann criterion allows to estimate
the melting temperature in a simple way.  However, the mechanism responsible
for the melting is still debated.  It was widely admitted
that the melting in three dimensions (3D)
occurs when one of the phonon modes is softened by the
temperature $T$ so that instability of the crystalline phase takes place
leading to the liquid phase. In 2D, this scenario is not valid. 
Mermin \cite{Mermin}
has shown in 1968 that long-range crystalline order is destroyed by $T$
if the elastic interaction is power-law-decayed with distance. Nelson and
Halperin\cite{Nelson} have shown that the
 2D melting is due to defects, analogous
to the case of Kosterlitz-Thouless transition for XY spins in 2D
\cite{kosterlitz73,kosterlitz74}.
 Inspired by this defect-mediated melting, several workers have attempted
to prove that in 3D the melting may be also due to dislocations and defects
\cite{Kleinert,Clements,Brostow,Gomez97,Kats,Bura}.
The soft-mode scenario, though theoretically possible,
is not the one frequently
observed in numerical calculations.  Works on defect theory up to
1989 have been summarized and developed  in the book by
Kleinert\cite{Kleinert}.  Recently, several numerical
investigations\cite{Clements,Brostow,Gomez97} and
analytic approximations\cite{Kats,Bura} have shown that
dislocations and defects are responsible for melting.
In particular, Burakowsky, Prestoni and  Silbar
\cite{Bura} have shown that melting properties of most of elements of the
Periodic Table can be explained by excitations of linear dislocations
in the crystal using a polymer theory.

Katsnelson and  Trefilov \cite{Kats} have also recently 
developed a route toward a melting theory
based in defects. They have emphasized the concept of the geometrical 
frustration \cite{Sadoc},
 i. e. the 3D euclidean space cannot be 
filled by the closest packing structure, the tetrahedral one. 
Therefore they proposed that superdense regions and voids around them are
to be formed near melting by thermal activation. These regions were
suggested to be
the precursor of the destruction of the crystalline order. 

In this paper, we investigate by Monte Carlo (MC)
simulation the melting of a 3D crystal
where atoms interact with each other via
the Lennard-Jones (LJ) potential.  We show that defects which 
occur in the solid phase when
the transition temperature is approached from below
are responsible for the melting.
Section II is devoted to a description of our model and MC method.
Results are shown in Section III. Concluding remarks are given in Section IV.

\section{Model and Monte Carlo Method}
      We consider a crystal of face-centered cubic (fcc) structure  which is
described by the following Hamiltonian
\begin{equation}
{\cal H}=\sum_{(ij)} U(r_{ij}) \label{equ.hami}
\end{equation}
where
the interaction between atoms at $r_i$ and $r_j$ 
is described by the potential
$U(r_{ij})$. For simplicity,
the distance dependence is supposed to be given by
the LJ potential
$U(r_{ij})= 4 \epsilon [(r_{0}/r_{ij})^{12}- (r_{0}/r_{ij})^{6}]$
where  $r_{ij}=r_{i}-r_{j}$, $r_{0}$ being a characteristic length 
of the system chosen in such a way that the NN distance in the fcc
lattice is equal to
$\sqrt2/2$ when only NN interactions are taken into account.
The fcc lattice constant is therefore equal to 1
in the ground state. 
Different potentials for cohesive interactions
are envisageable, for example the so-called Gupta many-body
potential\cite{Gomez95} which has been
recently used to study the melting process of a fcc crystal\cite{Gomez97}.
We have shown that defects created near melting play an important role in
the melting mechanism.
In the present paper we would like to clarify how defects are topologically
distributed and how they
destroy the crystal order. To this end, we choose the
LJ potential which is simpler for implementing a volume-variable
algorithm as will be described below. However, most of the conclusions
concerning the melting
mechanism are similar for these two potentials. Details of
a precise comparison will be given elsewhere.

In this work, we studied a fcc lattice with different system sizes
, using periodic boundary conditions,
at constant pressure.  
Interactions up to a cutoff distance $r_c=1.57$ have been taken into account.
This is about the fourth-nearest-neighbor distance in the perfect fcc crystal.

The following algorithm was used. Starting from the solid state where atoms
are on the fcc lattice sites, we heat the system up to a temperature $T$. We
equilibrate the system first locally at constant volume and then globally at
variable volume, as explained hereafter.  The local
equilibration at constant volume is done as follows:
we take an atom and  move it
to  a nearby random position. This position is accepted if it lowers
the atom energy. Otherwise it is accepted with a probability according to
the Metropolis algorithm. We repeat this for all atoms: we say
we achieve one MC step/atom.  Next, we
change the system volume by a random amount: all atom positions are thus
rescaled with the  volume variation. We recalculate the energy and 
accept or reject this new volume
using a constant pressure Metropolis algorithm.
For a general method, see Ref.\onlinecite{Allen}. In the following
we work at zero pressure.  
We find that the equilibrium is reached very fast with alternately 10
consecutive local MC steps/atom followed by one volume variation step,
and so forth.  In all, we performed about $10^5$ MC sweepings at each $T$.
Physical quantities such as averaged internal energy per atom $U$ and 
radial distribution function $g(r)$
are averaged over the next $10^5$ MC steps/atom.

From the plot of $\langle E\rangle$ versus $T$,
we can identify the transition temperature from solid to liquid state.
Furthermore, to investigate the melting mechanism we have computed
the following quantities: angular distribution of neighbors,
distribution of defects, number of nearest-neighbors (NN) etc.  
As we discuss below, due to the particular fcc structure we cannot use
the Voronoi method which would give confusing results if care is not taken.
 
\section{Results}

Let us show first in Fig. 1(a) $U$ versus $T$.
One observes a discontinuity of $U$
indicating clearly that the transition is of first-order as expected
for a 3D melting. 
This is confirmed by the jump of the average
fcc lattice constant
shown in Fig. 1(b). 
These figures show results for three different system sizes.
Since the transition is of first order,
the transition temperature $T_m$ cannot be defined with precision (due
to hysteresis). 
Moreover for sizes greater than 500 particles 
the melting temperatures fall  within the same interval. 
Therefore, in the following we will show results for the system of 
N=1372 particles.
We take $T_m$ as the lower limit
of the transition temperature region.  From Fig. 1, we take $T_m=0.76$.
Note that as our simulation cell does not contains surfaces, 
this transition temperature, corresponds to  
 the metastability limit of the crystal's ability to superheat 
rather than to the thermodynamical melting point \cite{Chokappa,Yip}.
In the following, we will call for simplicity, 
T$_m$ at this transition temperature and associate
the phenomena with an apparent melting point since 
the structure loss all local order at this temperature. 

Fig. 2 shows the radial distribution function
between atoms at a
low $T$ ($T=0.25$)  in the solid phase.
The first (second, third,...)
peak corresponds to the NN (NNN, third neighbors, ...) distance of the fcc
structure.

 To calculate the coordination number $c$ of a site, we integrate the radial
distribution function up to the first minimum.
For example in Fig. 2, integrating 
$g(r)$ over $r$ up to about $r_1=0.85$ (first minimum) will give $c=12$.
Integrating $g(r)$ from $r_1$ up to the next minimum at $r_2=1.1$ gives
$c_2=6$ and so on. 
 These coordination numbers verify the fcc structure.
We use this method instead of
the frequently used Voronoi construction because the Voronoi polyhedra
in a slightly distorted  fcc structure
have small faces and short edges  and the coordination
number could not be precisely  calculated  because these small faces
appear and disappear frequently due to 
thermal excitations \cite{Brostow,Richard}. 

Furthermore, to complete our structure determination, we also calculate
$g(\theta)$,  the angular distribution of neighbors around a site within
a plane up to a certain distance.
The function $g(\theta)$ gives the probability
to have two neighbors of an atom forming an angle $\theta$. 
We show in Fig. 3 an example of angular distribution taken at $T=0.25$ 
.  The peak at 60 (90, 120) degrees corresponds
to the angular distribution of the NN (NNN, 3rd NN) atoms 
in the fcc structure.

Let us increase $T$ near the transition.  The radial
distribution $g(r)$ at $T=0.7$
is shown in Fig. 4.
The main differences with the low $T$ structure shown in Fig. 2
are:
\begin{itemize}
\item All the minima between the peaks are raised. This is evidence   
that the crystal order is reduced at $T=0.7$. 
In particular, the minimum between the first
 and second peaks does not correspond to $g(r)=0$. This implies that
a diffusional dynamics is activated near melting. This is the origin of
the formation of defect clusters we will discuss below. 

\item Another feature is the strong reduction of the second and fourth
peaks. The cubic symmetry is therefore weakened and the 
remaining correlated part of the system is within the (111) planes.  
\end{itemize}

Integrating $g(r)$ over $r$ between
two consecutive minima, one
obtains the coordination number $c=12$ only at $74\%$ of the total
crystal sites. $15.5\%$ of the sites have 11 NN, $7.3\%$ have 13
NN and $2\%$ have 10 NN.  At this stage, it is worthwhile to stress that
though the temperature is still below $T_m$, such an important
percentage of sites ($26\%$)exhibit defects.
The crystal structure cannot be
therefore considered as a defect-free one which is used in the search
for a phonon soft-mode responsible for the elasticity instability at
the melting.

The angular distribution for $T=0.7$ is shown in Fig. 5. One
observes a strong deviation from the low-$T$ one shown in Fig. 3.
The cubic symmetry is reduced and the persistence of the (111) structure
is seen in this figure: the
peak at 90 degrees characteristic of the cubic symmetry
is broadened and reduced whereas the other two peaks of the (111) one
still have some structure.

 For  $T=0.79$, i.e. a 
temperature  just above the transition, we find the following striking result:
$30\%$ of sites with 11 NN, $25\%$ with 10 NN, $20\%$ with 12 NN,
$12\%$ with 9 NN, $4\%$ with 8 NN, $6\%$ with 13 NN, $1\%$ with 7 NN and
with 14 NN. 
Note that the percentage of 12-NN atoms has dropped from $74\%$ at $T=0.7$ to 
$20\%$ at this temperature. 
The average NN number is therefore 10.71.

Let us analyze now the structure of the observed defects. 
For this purpose we calculated
the radial and angular distributions {\it between defects} as follows:
whenever we are at a defect, i.e. atom with a coordination number
different from 12, we search for defects around it up to a certain
distance and realize histograms.
  Figure 6 shows the radial distribution
of defects at $T=0.7$ around defects with $c=13$.
One observes that i) the defects are surrounded
by other defects at a distance (0.696) slightly shorter than
the equilibrium NN distance indicating that at this  temperature
defects form clusters ii)there is a larger number of defects 
at a distance 0.876
very close to $\sqrt{3}/2$, indicating that defects occupy the sites at
the middle of the fcc cube which are normally vacant in the equilibrium
configuration.  This finding is important since it shows that defects forms
clusters with two shells: the inner are defects at 0.696 and the outer at
0.876 from the center defect.
At this temperature, this kind of defects are uncorrelated at larger
distance (see Fig. 6). The situation  for defects 
with $c=11$ is shown in Fig. 7. In this case the second peak is less intensive
and  order at larger distance appear as we discuss below.  
  Angular distributions between defects at $T=0.7$
shows a shift of a few degrees of the peaks toward
smaller angles for defects with $c=13$ as expected for $c$ larger than 12
(see Fig. 8).  However for defects with $c=11$, only the first
peak is lowered, while the other two peaks are shifted a few degrees higher.

Let us analyze more carefully the distribution of the
clusters formed by the defects.
We have computed the number of defects {\it inside} each cluster.
As we previously stated we have seen that defects are not isolated 
but they are arranged in separated inner and outer groups
within a defect cluster.
The outer groups appear and dissapear while the simulation evolves,
making the cluster size vary.
To quantify the sizes of these clusters we have obtained 
a histogram representing the probability to have a cluster 
with a given number of particles. The results are plotted  in
Figs. 9 at low, intermediate and near the melting point temperatures.
At low temperature [Fig. 9(a)], where the density of defects is very low, 
they appear mainly in pairs. These pairs of defects are completely isolated 
 and not important for the thermodynamical properties.
As the temperature increases, larger clusters of defects
start to be created [Fig. 9(b)].
Note that at temperatures near melting, clusters of all sizes are present
with the same probability up to a large given number of particles
[Fig. 9(c)].
When the cluster size reaches a given critical value the system melts.  
We can interpret these defect clusters as a set  of dislocation arrays or
alternatively as liquid zones inside a solid bulk. 
To enforce this interpretation
 we show the
radial distribution between defects in Fig. 10 for defects with $c=13$
at $T=0.79$.  One observes a large double peak indicating defects at
distances between NN and NNN equilibrium distances and pronounced
peaks at around 3rd neighbor distance ($\sqrt{2}$) and at 2.  This
suggests that defects are linked over large distances near the transition.
Note that this correlation is in fact insinuated below $T_m$ as is 
seen in Fig. 7 for defects of c=11 at T=0.7.  
The scenario proposed by Burakowsky,  Prestoni and  Silbar
\cite{Bura} is somewhat verified.
We recall that this  theory states that melting appears as a result of
dislocation generation. When the density of the dislocation array 
reach a given 
critical value, the entropy of these arrays compensates
the increase of energy
produced by the local breakdown of the perfect crystalline order.

Finally, we emphasize that our results were obtained with simulations
using a combination of local and volume updates.  However the use of
only collective updates with variable volume
as in our previous work\cite{Gomez97} does  not
alter our conclusion though the equilibrating time is somewhat longer.

\section{Concluding Remarks}
We have studied the melting mechanism of a fcc solid with the LJ potential.
The results show that defects created in the solid phase become
numerous enough to cause the crystal to melt. At the transition, about
one fourth of the total sites do not have the coordination number 12
of the perfect structure.  Just above the transition, only about 20$\%$
have 12 NN. Our analysis of the structure of defects shows that  a defect
is surrounded by atoms at $r$ about $\sqrt{2}/2$ (NN distance) and by atoms
at $r$ about $\sqrt{3}/2$ which is not the NNN distance (=1) of the perfect
structure. A closer examination shows that these atoms are themselves
defects created by dislocating atoms to somewhere between NN and NNN
distances. These dislocated positions correspond to 'bridge' positions
in the potential landscape.  We note that statistics taken between
defects shown above indicate that defects are linked together at
the transition. In other words, the assumption of linear defects by
Burakowsky, Prestoni and Silbar\cite{Bura} is somewhat verified here. 
By linear defects, one should understand 'strings' of defects which are not
necessarily straight lines of defects.  Of course, other aspects of their
theory should be further checked. But this is out of the scope of the
present work.

{\em Acknowledgements}: We are grateful to E.\ Jagla for useful
discussions. L.G. and A.D. thank the University of Cergy-Pontoise 
for hospitality.

'Laboratoire de Physique Th\'eorique et Mod\'elisation'
is associated with CNRS (ESA 8089).

\begin{figure}[htb]
\epsfig{file=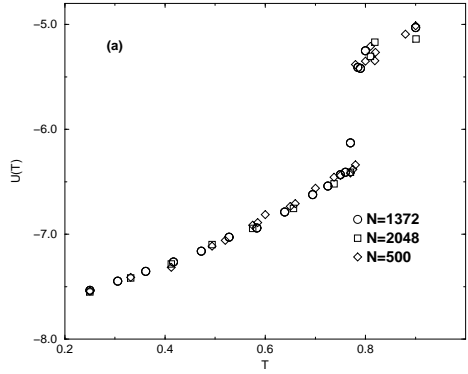,width=6cm,angle=-90}
\vskip .5 truecm
\epsfig{file=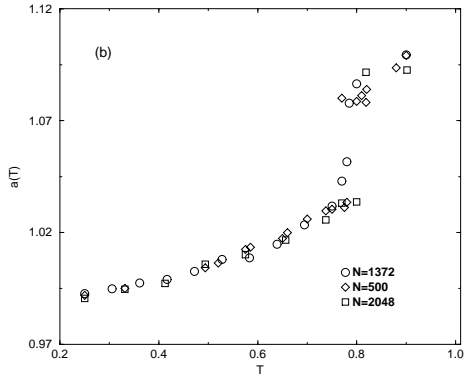,width=6cm,angle=-90}
\vskip .5 truecm
\caption{(a) Internal energy $U$ vs $T$   (b) lattice constant
vs $T$ for the system of N=1372. 
The jumps at the transition temperature indicate the
first-order character.}
\label{fig1}
\end{figure}

\begin{figure}[htb]
\epsfig{file=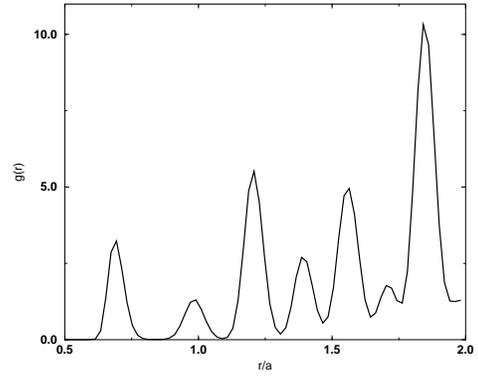,width=6cm,angle=-90}
\vskip .5 truecm
\caption{Radial distribution function g(r) at T=0.25.
This function gives the mean value of the number of particles sitting
at distance r from a given particle.} 
\label{fig2}
\end{figure}

\begin{figure}[htb]
\epsfig{file=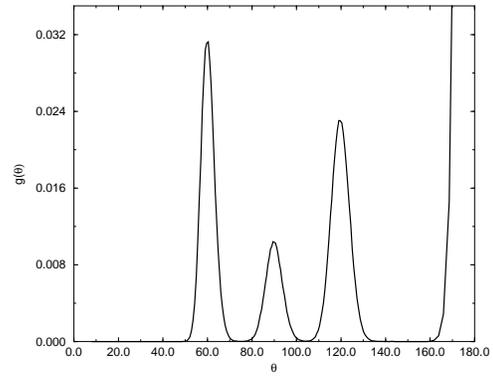,width=6cm,angle=-90}
\vskip .5 truecm
 \caption{
Angular distribution function as defined in the text at $T=0.25$.
The area below $g(\theta)$ is normalized to one.}
\label{fig3}
\end{figure}

\begin{figure}[htb]
\epsfig{file=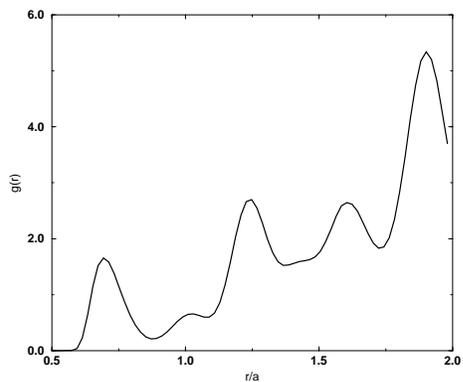,width=6cm,angle=-90}
\vskip .5 truecm
 \caption{
Function g(r) at $T=0.7$.}
\label{fig4}
\end{figure}

\begin{figure}[htb]
\epsfig{file=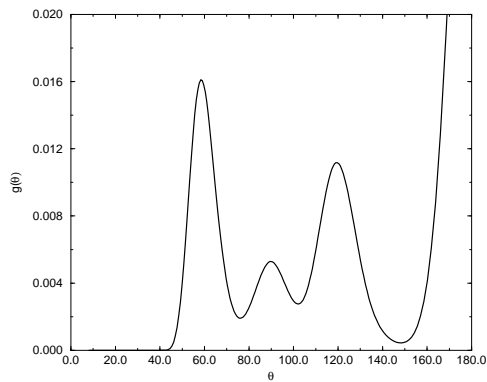,width=6cm,angle=-90}
\vskip .5 truecm
 \caption{
Angular distribution function g($\theta$) at $T=0.7$.}
\label{fig5}
\end{figure}

\begin{figure}[htb]
\epsfig{file=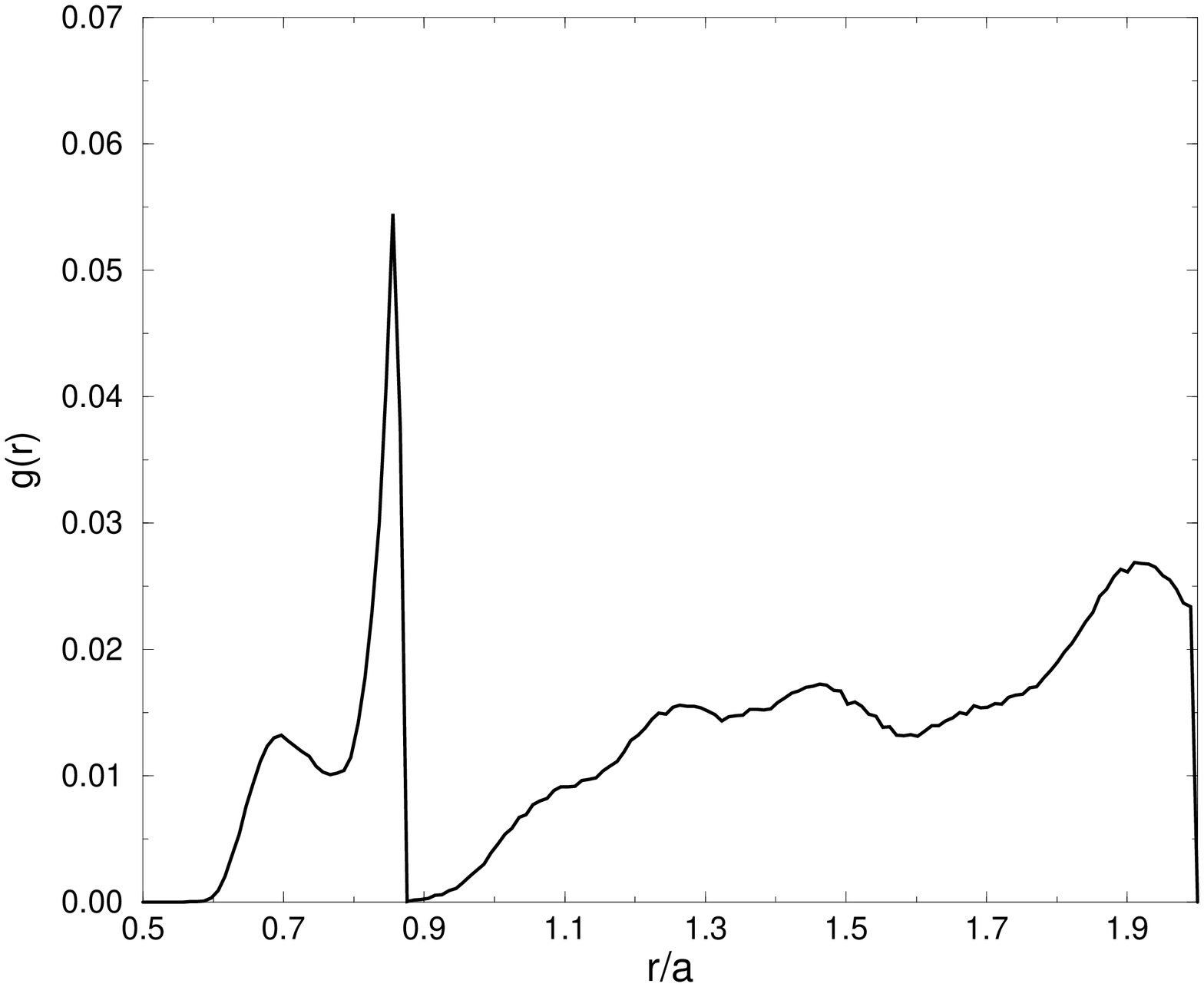,width=6cm,angle=-90}
\vskip .5 truecm
 \caption{
Radial distribution function {\it between} defects of coordination
c=13 at $T=0.7$. Note the different scale with Fig. 2 because
we fix the area
below $g(r)$ to be one here} 
\label{fig6}
\end{figure}

\begin{figure}[htb]
\epsfig{file=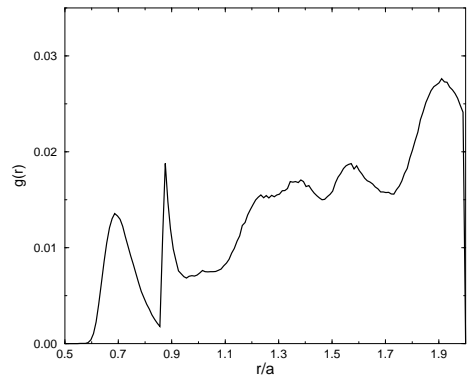,width=6cm,angle=-90}
\vskip .5 truecm
 \caption{
The same as fig. 6 but for defect with c=11.} 
\label{fig7}
\end{figure}

\begin{figure}[htb]
\epsfig{file=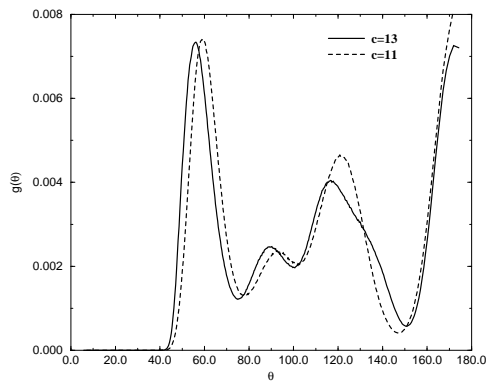,width=6cm,angle=-90}
\vskip .5 truecm
 \caption{
Angular distribution around defects of coordination c=11 and 13 at T=0.5}
\label{fig8}
\end{figure}

\begin{figure}[htb]
\epsfig{file=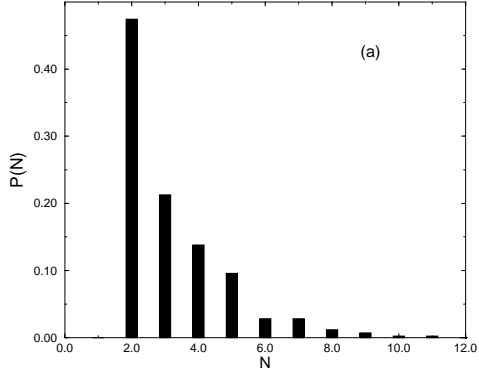,width=6cm,angle=-90}
\vskip .5 truecm
\epsfig{file=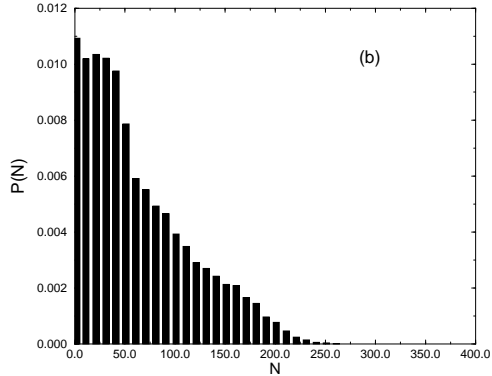,width=6cm,angle=-90}
\vskip .5 truecm
\epsfig{file=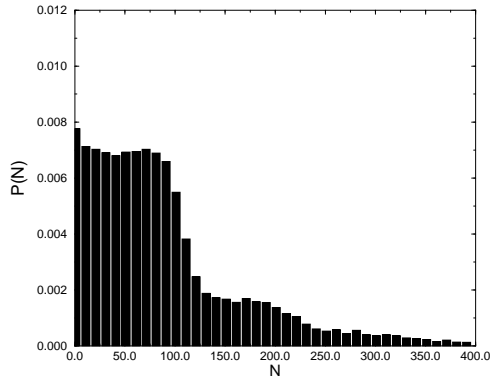,width=6cm,angle=-90}
\vskip .5 truecm
 \caption{
Histogram of the number of defect in a cluster for  a) T=0.3, b) T=0.58
and c) T=0.725 
} 
\label{fig9}
\end{figure}

\begin{figure}[htb]
\epsfig{file=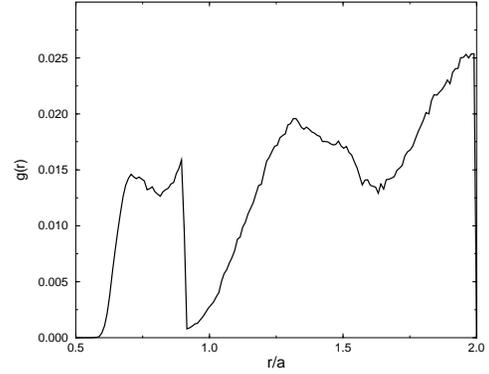,width=6cm,angle=-90}
\vskip .5 truecm
 \caption{
Radial distribution function {\it between} defects of coordination
c=13 at $T=0.79$ .} 
\label{fig10}
\end{figure}

%\begin{figure}[htb]
%\epsfig{file=fig10.eps,width=6cm,angle=-90}
%\vskip .5 truecm
% \caption{
%Radial distribution function {\it between} defects of coordination
%c=11 at $T=0.55$ for the system of N=500 particles.} 
%\label{fig10}
%\end{figure}

%\begin{figure}[htb]
%\epsfig{file=fig11a.eps,width=6cm,angle=-90}
%\epsfig{file=fig11b.eps,width=6cm,angle=-90}
%\vskip .5 truecm
% \caption{
%Radial distribution function for a temperature just above (solid line)
% at the melting (long dashed line)
% and bellow the melting temperature (dot-dashed line)
%for the system interacting via: (a) LJ 
%potential, (b) Many body potential.}
%\label{fig11}
%\end{figure}

\newpage

\end{document}